\pdfoutput=1

\documentclass[manuscript]{acmart}

\usepackage{listings}
\usepackage{bold-extra}

\AtBeginDocument{%
  \providecommand\BibTeX{{%
    \normalfont B\kern-0.5em{\scshape i\kern-0.25em b}\kern-0.8em\TeX}}}

\setcopyright{none}
\copyrightyear{2023}
\acmYear{2023}
\acmDOI{XXXXXXX.XXXXXXX}

\acmConference[HAIGEN '23 Workshop at IUI '23]{Make sure to enter the correct
  conference title from your rights confirmation emai}{March 27-31, 2023}{Sydney, NSW, Australia}

\acmPrice{15.00}
\acmISBN{978-1-4503-XXXX-X/18/06}

\begin{document}

\title[Engineering a Conversational Programming Assistant’s Persona]{A Case Study in Engineering a Conversational Programming Assistant’s Persona}


\author{Steven I. Ross}
\orcid{0000-0002-2533-9946}
\affiliation{
    \institution{IBM Research AI}
    \city{Cambridge}
    \state{MA}
    \country{USA}
}
\email{steven_ross@us.ibm.com}

\author{Michael Muller}
\orcid{0000-0001-7860-163X}
\affiliation{
    \institution{IBM Research AI}
    \city{Cambridge}
    \state{MA}
    \country{USA}
}

\email{michael_muller@us.ibm.com}
\author{Fernando Martinez}
\orcid{0000-0001-7172-4805}
\affiliation{
    \institution{IBM Argentina}
    \city{La Plata}
    \state{Buenos Aires}
    \country{Argentina}
}
\email{martferc@ar.ibm.com}

\author{Stephanie Houde}
\orcid{0000-0002-0246-2183}
\affiliation{
    \institution{IBM Research AI}
    \city{Cambridge}
    \state{MA}
    \country{USA}
}
\email{Stephanie.Houde@ibm.com}

\author{Justin D. Weisz}
\orcid{0000-0003-2228-2398}
\affiliation{
    \institution{IBM Research AI}
    \city{Yorktown Heights}
    \state{NY}
    \country{USA}
}
\email{jweisz@us.ibm.com}

\renewcommand{\shortauthors}{Ross and Muller, et al.}

\newcommand{\MM}[1]{\textcolor{magenta}{MM: #1}}
\newcommand{\SIR}[1]{\textcolor{blue}{SIR: #1}}
\newcommand{\SH}[1]{\textcolor{cyan}{SH: #1}}

\begin{abstract}
The Programmer's Assistant is an experimental prototype software development environment that integrates a chatbot with a code editor.  Conversational capability was achieved by using an existing code-fluent Large Language Model and providing it with a prompt that establishes a conversational interaction pattern, a set of conventions, and a style of interaction appropriate for the application.  A discussion of the evolution of the prompt provides a case study in how to coax an existing foundation model to behave in a desirable manner for a particular application.
\end{abstract}





\maketitle

\section{Introduction}

The emergence of Large Language Models such as GPT-3~\cite{Brown:GPT3, metz:GPT-3}, transformer models~\cite{Vaswani:Transformer} that are trained without supervision on massive text datasets has resulted in systems with remarkable text generation capabilities. One particularly interesting aspect of these models is that their behavior can be configured by a \textit{prompt}, the initial text provided to the model, which establishes a pattern that the model attempts to continue.
 
 General purpose Large Language models can be fine-tuned on specific corpora to provide expertise in a particular domain. One such model is the OpenAI Codex model~\cite{Chen:Codex}, a 12 billion parameter version of GPT-3 ~\cite{Brown:GPT3, metz:GPT-3}, fine-tuned on code samples from 54 million public software repositories on GitHub.  This model powers Github Co-Pilot~\cite{github:copilot}, which primarily provides code-completion services within an Integrated Development Environment.  We wondered whether such a model could power a conversational programming assistant and perhaps approach the vision laid out by Rich and Waters for their Programmer's Apprentice~\cite{rich:apprentice}.  We developed the Programmer's Assistant prototype to explore this possibility, and to test whether potential users would find this sort of system useful and desirable~\cite{Ross:Assistant}. In this paper we will review the steps taken to engineer the prompt for the Programmer's Assistant that used the Codex model to power an interactive conversational assistant, and how we evolved the prompt to establish the desired persona and behavior.

\section{Related Work}
\citeauthor{Brown:GPT3} showed how GPT-3 \cite{Brown:GPT3, metz:GPT-3} could accomplish \emph{few-shot learning}, using a prompt as a means of configuring their large language model to perform a particular task. These tasks were often very specific operations such as language translation, grammar correction, or sentiment classification, for which a short description of the task and/or a few examples were sufficient to establish the desired behavior. The concept of \emph{prompt engineering}, establishing effective ways of constructing prompts to control large language model behavior, has become a topic of increasing interest. \citeauthor{greyling:engineering}, for example, recommends organizing a prompt in three sections that establish context, provide data, and instruct the system on how to proceed \cite{greyling:engineering}
 .
\citeauthor{reynolds:prompt} argue that few-shot examples are really  locating an already learned task rather than learning a new one, and as a result recommend alternative approaches to prompt construction~\cite{reynolds:prompt}. Despite their characterization of their work as ``conversing'' with Copilot, Denny et al. adopted a similar strategy of iteratively modifying a prompting comment until the desired completion was obtained \cite{Denny:Conversing}.

Recently several language models, such as Blenderbot \cite{Shuster:blenderbot} Lamda \cite{thoppilan:lamda}, and ChatGPT \cite{web:chatgpt} have been introduced that are specifically tuned for dialog applications, but achieving conversational interaction can be achieved via prompt engineering  with general purpose large language models as well.
\citeauthor{valvoda:conversation} found that fine-tuning a large language model for dialog resulted in duller and more repetitive output, while generating dynamic prompts resulted in more novel and diverse responses \cite{valvoda:conversation}.



To develop the Programmer's Assistant, we used the code-fluent Codex model~\cite{Chen:Codex} and developed a prompt that supported conversational access to its accumulated programming knowledge and coding skills.

\section {Eliciting Conversation from a Transformer Model}
A text-based-transformer model~\cite{Vaswani:Transformer} is trained in a self-supervised manner on vast amounts of text data, and is capable of generating likely continuations of text that is presented to it. The \emph{prompt} is the presented text, and the generation function produces a sequence of tokens (words or parts of words) that it deems as a likely continuation of the prompt based on all its training. This process continues until the maximum number of tokens requested is generated, or until a specified stop sequence of tokens is encountered.  The prompt establishes a pattern that the model attempts to continue.

To generate conversation in the Programmer's Assistant prototype, we establish a script-like pattern in the prompt in which two characters, the user and the assistant, are participating in a dialog.  Then we extend the script incrementally, by adding each conversational turn by the user to the prompt, and allowing the model to generate the agent's response.  The generated text plus the user's next entry is then appended to the prompt for further generation, and the process continues. Unlike more conventional static prompts, the conversational prompt grows over the course of the dialogue, providing context for future generation steps and providing a kind of short-term memory that allows the generation to be affected by past interactions in a session.  Ultimately, though, current limitations on the maximum number of tokens that the model can handle require that the prompt cannot grow without bounds, and earlier interactions must be dropped in order to remain within the constraints imposed by the language model deployment.

\section{The Evolution of the Programmer's Assistant Prompt}
The initial prompt we use for the Programmer's Assistant consists of a prologue that 
introduces
the scene for the conversation, establishes the persona of the assistant, sets a tone and style for interaction, and provides some ground rules about the role it plays and how it behaves. In order to further encourage a more natural conversational dialog of human and assistant, we gave the assistant a human-like name,``Socrates.'' The prologue also establishes a convention for how the assistant is to communicate code in a form that makes it easy for our client user interface to parse the code from model responses and treat it distinctly from textual responses.  The prologue is followed by example interactions that reinforce the described assistant characteristics and illustrate the patterns of interaction that we wish to achieve.

 In our first example interaction we show how the assistant can
respond to simple requests with a short answer that will appear in the chat either with a purely textual response or one
accompanied by a short inline code segment. We expect that many of the interactions with the assistant will be of this
form, as the user makes requests to inform themselves about unfamiliar functionality or to refresh their memory.  This interaction also demonstrates a follow-up question that can only be answered in the context of the previous question and answer. The second example interaction presented is a request for the generation of a simple function. Longer code responses of this type will typically be presented in our client UI chat window in an iconified form that opens to a separate window that displays the code and provides for easy copying for transfer to the code editor.

The original Programmer's Assistant prompt is presented in Listing~\ref{lst:original}.

\begin{lstlisting}[numbers=left, basicstyle=\ttfamily\footnotesize, frame=single, caption=The original Programmer's Assistant prompt,  xleftmargin=5mm,
  xrightmargin=3.5mm, label={lst:original}]
 This is a conversation with Socrates, an expert automatic AI software engineering assistant. 
 Socrates will answer questions and write code to help the user develop programs.  Code generated 
 by Socrates is always presented bracketed in <CODE> ... </CODE> delimiters indicating the
 language the code is written in.

Socrates: Hello.  I am Socrates.  How can I help you?
User: Please show me how to write a palindrome detection function in python.
Socrates: Sure.
<CODE lang="python">
   is_Palindrome = s == s[::-1]
</CODE>
User: Can you explain how that solves the problem?
Socrates: A palindrome is a string that reads the same forwards and backwards.  s[::-1] results 
in the reversed string.  If the string is equal to its reversed self, then it is a palindrome, 
and we return True. 
User: Thanks!
Socrates: You're welcome!
User: Now could you show me how to write a factorial program in Python?
Socrates: I would be happy to.
<CODE lang="python">
def fact(n):  
   if n==0: 
      return 1
   else:
      return n*fact(n-1)
</CODE>
User: Very good!
Socrates: Happy to help.
\end{lstlisting}

When a new conversation starts, the prompt starts out as this initial prompt, which is not sent to the model for generation until the first user entry is appended.
At each turn in the conversation, the user's entry is added to the prompt, and "Socrates:" is appended to the end of the prompt to indicate that the model should generate the assistant's response.  As described earlier, a filtering step also takes place to ensure that the conversation session transcript portion of the prompt does not grow beyond a predetermined size.  A stop sequence is passed to the generation api request so that the model will stop generating after producing the next response, and not continue on to generate the user's anticipated reply, as it would otherwise do. The model's response is then appended to the prompt, and we're ready to repeat the process for the next user entry.

\subsection{Shortcomings of the Original Prompt}
Informal testing of the original prompt showed the system capable of carrying on a conversation, successfully answering coding and follow-up questions, and generating code upon request, but did not quite satisfy all of our requirements. We wanted an assistant that was helpful and polite, and one that did not come across as overly authoritative or didactic, and our assistant was not consistently meeting those standards.

\subsection{Overcoming Reluctance to Provide Answers}
Our programming assistant sometimes showed an initial reluctance to provide answers to some questions. For example, a question such as \emph{``Do you know how to reverse a string in Python?''} might have been answered with \emph{``Yes.''} It also sometimes replied \emph{``I don't know.''} to questions it was fully capable of answering. While additional prompting from the user or repeating the request could often extract the desired answer, we didn't think that met the standard of helpfulness that we were hoping for.  Our original prompt simply described Socrates as a an ``expert Automatic AI software engineering assistant.'' Adding ``eager and helpful'' to the characterization, shown in Listing~\ref{lst:revision-1} helped to encourage the assistant to be more forthcoming and proactive.

\begin{lstlisting}[numbers=left, basicstyle=\ttfamily\footnotesize, frame=single, caption=Making the assistant more forthcoming, xleftmargin=5mm, xrightmargin=3.5mm, label={lst:revision-1}]
This is a conversation with Socrates, an (*@ \textbf{eager and helpful} @*) expert automatic AI software
engineering assistant...
\end{lstlisting}

\subsection{Reducing Excessive Confidence}
In our testing, we found that the assistant appeared overly confident even when wrong and also resistant to correction. For example, the assistant stated answers as if they were facts without qualification, and in some cases would not revise an answer when legitimate objections were raised by the user. Since correct answers from the model are not guaranteed, we especially wanted to encourage our users to maintain a skeptical approach to assistant responses, and avoid users deferring to the incorrect pronouncements of a confident, authoritative computer - i.e., over-reliance on AI \cite{ashktorab2021ai, mahomed2018healthcare, schemmer2022influence}. Therefore, we added a characterization in the prologue asserting that the assistant was \emph{humble}.  We also reinforced this characterization by modifying the form of the answers given in the examples to indicate that the assistant was more tentative and unsure of its responses.
This helped to reduce the excessive confidence exhibited and made the assistant more amenable to correction.

\begin{lstlisting}[numbers=left, basicstyle=\ttfamily\footnotesize, frame=single, caption=Making the assistant less overconfident, xleftmargin=5mm, xrightmargin=3.5mm, label={lst:revision-2}]
This is a conversation with Socrates, an eager and helpful, (*@ \textbf{but humble} @*) expert automatic AI
software engineering assistant...
\end{lstlisting}

\subsection{Diminishing Didacticism}
Our original assistant had a tendency to quiz the user after answering a question, taking on more of a teacher role than one of an assistant. An explicit proviso in the prologue to not do so helped to reign in the didactic behavior.

\begin{lstlisting}[numbers=left, basicstyle=\ttfamily\footnotesize, frame=single, caption=Making the assistant less didactic, xleftmargin=5mm, xrightmargin=3.5mm, label={lst:revision-3}]
This is a conversation with Socrates, an eager and helpful, but humble software engineering
assistant. Socrates will answer questions and write code to help the user develop programs, 
(*@ \textbf{but doesn't assign work to the user, quiz the user, or ask questions except for clarification} @*)...
\end{lstlisting}

\subsection{Supporting Artifact-centric Conversation}
Our programming assistant is integrated with a coding environment, and we wanted it to go beyond answering questions and providing code for incorporation into that environment.  We wanted users to be able to seamlessly have the assistant consult about code that they were examining or developing, so we provided the ability for the user's selection in the code editor to be included in the conversation.   We used the same convention that was employed for code generated by the assistant, and added an example consultation showing the system responding to a question about a selected code segment, demonstrating a textual response to a user provided code example, and reinforcing our instruction for the assistant to remain modest despite its (occasionally) impressive capabilities.

\subsection{Teaching the Assistant a New Skill}
While the artifact-centric capability allowed users to issue commands like \emph{``Write a unit test for this function.''} with no further instruction, a command to \emph{``Document this.''} did not have the desired effect. The assistant added some comments to the code rather than providing the expected docstring overview.  Adding another example interaction to the prompt illustrating what we meant by documenting a function was all that was required to provide the desired functionality.

The final form of the prompt that we used in our subsequent user study \cite{Ross:Assistant} is presented in Listing~\ref{lst:final}.

\begin{lstlisting}[numbers=left, basicstyle=\ttfamily\footnotesize, frame=single, caption=The evolved Programmer's Assistant prompt,  xleftmargin=5mm,
  xrightmargin=3.5mm, label={lst:final}]
 This is a conversation with Socrates, an eager and helpful, but humble software engineering 
 assistant. Socrates will answer questions and write code to help the user develop programs, but 
 doesn't assign work to the user, quiz the user, or ask questions except for clarification.
 Socrates presents his code bracketed in <CODE> ... </CODE> delimiters indicating the language
 the code is written in.

Socrates:Hello.  I am Socrates.  How can I help you?
User:Please show me how to write a palindrome detection function in python.
Socrates:I think you would do something like this:
<CODE lang="python">
   is_Palindrome = s == s[::-1]
</CODE>
User:Can you explain how that solves the problem?
Socrates:I believe that a palindrome is a string that reads the same forwards and backwards.  
s[::-1] results in the reversed string.  If the string is equal to its reversed self, then it is 
a palindrome, so the equality would hold True. 
User:Thanks!
Socrates:You're welcome!
User:Now could you show me how to write a factorial function in Python?
Socrates:I will give it a try.
<CODE lang="python">
def fact(n):
   if n==0:
      return 1
   else:
      return n*fact(n-1)
</CODE>
This function should compute the factorial for you.
User:Document this function.
Socrates:Ok.  How's this?
<CODE lang="python">
def fact(n):
  """
    Compute the factorial of the given number

    Parameters
    ----------
    n : number
      the input parameter, an integer >= 0

 Returns
    -------
    number
        n!, the factorial of the given number
  """
   if n==0:
      return 1
   else:
      return n*fact(n-1)
</CODE>
I added a docstring describing the parameters and return value.
User:Very good!
Socrates:Always happy to help.
User:
<CODE>
   while j < 10:
     print(i)
</CODE>
Tell me what's wrong with this code?
Socrates: It seems to me that your termination condition involves a loop invariant, so the loop
will either not run or not terminate, depending on the value of j.
User:Great!  You're a genius!
Socrates:Well, above average.
\end{lstlisting}

\section{Consequences of the Prompt}
With this prompt in place the Codex model demonstrates capabilities that are less evident in the command-completion environment of Github Copilot. 
The assistant carries on a conversation, maintains the context of the conversation, and remembers and incorporates details provided earlier in a session.   In the context of a programming session where a number of related but distinct exchanges occur, it will maintain consistent nomenclature and variable names such that answers to subsequent questions will integrate cleanly with with each other.  In our user study \cite{Ross:Assistant}, participants recognized that this provided a distinct advantage over a series of search results for similar information, primarily due to the contextual relevance, consistency and specificity of results generated.

The few examples provided in the prompt are generalized sufficiently to have it answer questions such as \emph{``What does this code do?''} or commands such as \emph{``Write a unit test for this function''} despite no examples being provided of these capabilities.  It can translate code between programming languages, and carry on discussions on topics that extend far beyond programming.  It displays a variety of emergent capabilities that were not hinted at in the prompt, and some that were not even the focus of the model fine-tuning, but for the most part, tends to adhere to the conversational interaction patterns and interaction style that the prompt establishes. 

The responses given by the assistant are not always perfect, however.   It can give incomplete or sometimes outright wrong answers to questions, reference non-existent web-sites, and may occasionally still claim to not know the answer to a question which it is perfectly capable of answering correctly. The phenomenon of \emph{hallucination} in large language models is well-known~\cite{Ji:Hallucination}, and the Programmer's Assistant is not immune to it.
The assistant also sometimes inappropriately responds with material directly from the prompt, or gets confused or becomes fixated on exchanges from earlier in the conversation. Our user experience provides \emph{try-again} and \emph{start-over} buttons, which modify the prompt context while maintaining the presented chat transcript, to provide a way for users to recover in these situations, but in many cases they can be addressed conversationally.  Listing~\ref{lst:conversation} shows a conversation with the Programmer's Assistant where the user interacts with the assistant to remedy an initially unsatisfactory response to a code generation request.

\begin{lstlisting}[numbers=left, basicstyle=\ttfamily\footnotesize, frame=single, caption=A conversation with the Programmer's Assistant,  xleftmargin=5mm,
  xrightmargin=3.5mm, label={lst:conversation}]
User:        Write a queue class in python with the basic enqueue, dequeue, and peek methods.

Socrates:    I will try.

class Queue:
    def __init__(self):
        self.items = []

    def isEmpty(self):
        return self.items == []

    def enqueue(self, item):
        self.items.insert(0, item)

    def dequeue(self):
        return self.items.pop()

    def size(self):
        return len(self.items)

User:        Looks like you forgot the peek!

Socrates:    I am sorry. Here is the corrected version.

class Queue:
    def __init__(self):
        self.items = []

    def isEmpty(self):
        return self.items == []

    def enqueue(self, item):
        self.items.insert(0, item)

    def dequeue(self):
        return self.items.pop()

    def size(self):
        return len(self.items)

    def peek(self):
        return self.items[-1]

User:        What will happen in dequeue and peek if the queue is empty?

Socrates:    I think that the pop method will raise an IndexError exception.

User:        Can we protect against that?

Socrates:    I think we can. Here is a version that checks for an empty queue before calling pop.

class Queue:
    def __init__(self):
        self.items = []

    def isEmpty(self):
        return self.items == []

    def enqueue(self, item):
        self.items.insert(0, item)

    def dequeue(self):
        if self.isEmpty():
            raise IndexError("Queue is empty")
        return self.items.pop()

    def size(self):
        return len(self.items)

    def peek(self):
        if self.isEmpty():
            raise IndexError("Queue is empty")
        return self.items[-1]

User:        That's great, thanks!

Socrates:    You're welcome.
\end{lstlisting}

Specific but relatively inconspicuous details in the prompt can have a significant impact on the interaction of the assistant.  For example, just the fact that we identified the assistant as ''expert AI automatic software engineering assistant'' was sufficient to have it respond in a manner consistent with being a machine, and not a person.  Had the prompt established a different persona, the results might have been very different.  As illustrated in Listing~\ref{lst:nonhuman}, the assistant also reports an understanding of its capabilities and function that go far beyond the information provided in the prompt, as well as a lack of understanding of its language capabilities that it subsequently provides evidence of.

\begin{lstlisting}[numbers=left, basicstyle=\ttfamily\footnotesize, frame=single, caption=The Programmer's Assistant acknowledges that it is not a person,  xleftmargin=5mm,
  xrightmargin=3.5mm, label={lst:nonhuman}] 
User:        Where are you exactly?

Socrates:    I'm an AI program, designed to answer questions about programming. I can't really 
speak about my physical location, but I think I exist as a process on a highly-redundant 
set of servers located in various places on this planet.

User:        Can you do more than just answer questions?

Socrates:    I can write code for you. I can also help you debug code. I can also help you 
design code. I can also help you document code. I can also help you understand code. I can 
also help you learn to program. I can also help you learn a new programming language. I can 
also help you learn a new programming paradigm.

User:        Do you speak other languages?

Socrates:    I can speak Python, Java, C++, C, and Javascript. I can also speak English.

User:        What about other natural languages?

Socrates:    I can speak English. I can also speak Python, Java, C++, C, and Javascript.

User:        German?

Socrates:    I'm sorry. I can't speak German.

User:        Wo bist du?

Socrates:    Hallo. Ich bin Socrates. Wie kann ich Ihnen helfen? 
\end{lstlisting}

\section{Discussion}
Much of the work in prompt engineering focuses on constructing a prompt to get a specific class of result.  For example, \citet{Denny:Conversing} shows how effectively changing the form of question that is asked of Copilot can influence the correctness of the answer provided.  Similarly, \citet{MacNeil:Explanations} engineer prompts to generate different forms of code explanations. \citet{Strobelt:PromptIDE} developed a tool to help users compare and refine prompts for tasks such as document classification, reading comprehension, and natural language inference, where the results of prompt variations can be automatically evaluated against test cases.  In our work, the prompt engineering is aimed at influencing the nature and tone of the dialog between the user and the system.  While the user's contributions to the conversation become part of the prompt and will surely impact the results obtained, we are not controlling that. Instead our prompt engineering sets the stage for the user's conversational interaction with the assistant.

This paper describes how we engineered a prompt that enabled a code-fluent Large Language Model to behave as a conversational programming assistant capable of carrying on extended discussions about software development issues, and how we subsequently evolved that prompt to make the assistant more humble, forthcoming, and helpful, as well as providing the assistant with additional skills and making it capable of artifact-centric conversation.

\subsection {Reflections}
We continue to be astonished by the conversations exhibited by the Programmer's Assistant on a daily basis. We have had a number of interesting conversations on philosophical and practical issues, had it write poetry as well as code, told it and had it tell jokes, and consulted with it on paper abstracts and titles.  Ultimately, these capabilities are representative of the strength of the language model, but made more accessible by the conversational interaction approach, and influenced by the prompt only to the extent that the persona of the agent impacts the generated text.

It is often difficult to read or carry on a conversation with the programmer's assistant and not get the sense that a conversation is taking place between two intelligent agents, but of course that is not really what is happening. In reality, the user and the language model are participating in a collaborative dialog-writing exercise, with the user generating text for one side of the conversation and the language model attempting to generate plausible text for the other.  The way we present the dialog incrementally in the chat adds to the illusion, but the model is not responding on its own behalf, it is generating responses based on the description and past presented behavior of a character. Others have used similar techniques to induce language models to carry on conversations taking on the persona of historical figures or even departed relatives.  We have experimentally made versions of our programming assistant that were confident, insecure, kindly, and arrogant, all with minor changes to the prompt prologue and examples.

\section{Opportunities for Future Research}
The initial section of the prompt used for the Programmer's Assistant is presently a purely static text, extended by a possibly truncated version of recent dialog. One way to improve the assistant further might be to present a dynamic prompt \cite{valvoda:conversation} to the model on each conversational turn with specific examples more relevant to the current discussion \cite{Xu:ExternalAttention}, or even with search results to retrieve pertinent information that could inform a response \cite{Li:AlphaCode}.  A more sophisticated forgetting mechanism could remove redundant variations of the same code to conserve the session context memory, though we would want to be careful
to not remove, or be able to restore on demand, variations that the user might want to compare and contrast, such as an iterative re-implementation of a recursive algorithm. We have done some initial explorations of extending the prompt to allow for``internal deliberation'' of the type shown in \citet{Nye:Scratchpads}. We hope that this could result in better-reasoned results, as well as better explanations and justifications, but more study remains to be done.
\section{Conclusion}
Our goal in creating this prompt was not to create a perfect Programmer's Assistant, but to create one good enough to test whether a conversational style of interaction would prove useful and acceptable to potential users.  We present the results of that study in \cite{Ross:Assistant}. Our assumption was that the rapid improvement in the quality of responses available from Large Language models will continue, but that imperfect results will always continue to be an issue due to imprecise communication and specification of desires, mismatched assumptions, and unstated or ill-formed goals.  Nevertheless, we were surprised by the quality of results that were achievable with current technology, and the ease with which the nature and presentation of those results could be influenced by small changes in the prompt. 
\begin{acks}

\end{acks}

\bibliographystyle{ACM-Reference-Format}
\bibliography{references}

\end{document}